\documentstyle[12pt]{article}

\begin{document}
\title{Duality, generalized Chern-Simons terms and gauge transformations in a high-dimensional curved spacetime}

\author{Hisashi Echigoya, Tadashi Miyazaki and Tomohiko Shibuya \\
{\it Department of Physics, Science University of Tokyo}, \\
{\it Kagurazaka, Shinjyuku-ku, Tokyo 162-8601, Japan}}

\date{}

\maketitle
\vspace{10mm}

\begin{abstract}
With two typical parent actions we have two kinds of dual worlds: 
i) one of which contains an electric as well as magnetic current, 
and ii) the other contains (generalized) Chern-Simons terms.
All these fields are defined on a curved spacetime of arbitrary (odd) 
dimensions. A new form of gauge transformations is introduced and 
plays an essential role in defining the interaction with a magnetic 
monopole or in defining the generalized Chern-Simons terms.\\

\noindent
PACS number(s): 02.40.Ky, 11.10.Kk, 11.15.-q
\end{abstract}
\setlength{\baselineskip}{7.8mm}

Theory of duality, which represents physical contents with two different 
sets of fields, has recently been playing an important role in particle 
physics.
Among the various works along this line are the self-dual massive gauge 
theory \cite{Townsend} and the topological massive gauge theory \cite{Deser},
and it was shown that both theories are {\it dual} to each other.
More recently, P.K.Tripathy and A.Khare replaced Deser and Jackiw's Lagrangian 
of topological massive gauge fields by the nonlinear Born-Infeld Lagrangian 
\cite{Tripathy,Born}, which had been proposed in order to improve 
the short-distance divergence difficulities \cite{Born}.
The Born-Infeld Lagrangian, by the way, has also become a tool to describe the
dynamics of D-brane gauge fields \cite{Tseytlin}. \\

To definitely realize the idea of duality one first introduces some fields 
with a {\it parent action}.
Then, one calculates the equations of motion, through the principle of 
variation, for one or two particular fields. 
The use of the equations of motion allows us to eliminate the fields in the 
parent action, and we have an action as a functional of the leftover fields.
On the other hand we {\it first} write down the equations of motion for 
the latter leftover fields based on the parent action, and, eliminating 
them from the action with these equations of motions, we obtain an action 
for the fields which were eliminated in the first step.
In this way we are to have two kinds of actions --- functionals for the 
two kinds of fields.These two actions are considered to be {\it dual} 
to each other, and the contents which are represented by each action are 
{\it dual physical worlds}. This method becomes a very useful tool to 
define duality and was described neatly in Ref.\cite{Lindstrom}.
If one further goes into the dual theory of interacting vector fields with 
matter with U(1) symmetry, one must add a Thirring-like interaction 
to the Lagrangian. \\

The discussions mentioned above are either restricted to the spacetime of 
dimensions $3=1+2$ or to the {\it flat} Euclidean (or Minkowski) space.
Or, one was faced with the restricted form of gauge fields.
How can we formulate the dual theory if we attack the problem as generally as 
possible ?
The space may be {\it curved}; 
the dimensions may be arbitraily large (but maybe odd);
the gauge fields may be any antisymmetric tensors of high ranks.
How can we set up the formulation with this new facet of situations and show
the dual relation between actions or fields?
Our main purpose in this paper is to answer these questions.
In the process we will find that two types of gauge transformations appear, 
one of which plays a role in defining a magnetic monopole current. \\

We follow technical treatment of Ref.\cite{Miyazaki}, where the present 
authors developed, quite mathematically, theory of fields with 
an electromagnetic current as well as a magnetic monopole current 
in the curved spacetime of arbitraty dimensions.
There, what describes the physical world are $q$-forms ($q$ 
dimensional differential forms) over $d$ dimensional spacetime.
We will not write down differerntial forms {\it in components} as far as
it be unnecessay, because such caluculations lead us to boresome work
in case of the spacetime being curved \cite{Miyazaki}.
Now, let us start with the following parent action.
\begin{eqnarray}
S_{p}[F^{(q)},A^{(q-1)},B^{(q+1)}]&=&\frac{1}{2}(F^{(q)} , F^{(q)})
                                     -(F^{(q)} , d A^{(q-1)})
                                     -(F^{(q)} , \delta B^{(q+1)}) \nonumber \\
                                   & &-(F^{(q)} , G^{(q)})
                                      +(A^{(q-1)} , J^{(q-1)})
                                      +(B^{(q+1)} , K^{(q+1)}) \nonumber \\
                                   & &+\frac{m^2}{2}(A^{(q-1)} , A^{(q-1)})
                                      +\frac{{\mu}^{2}}{2}(B^{(q+1)} , B^{(q+1)}) ,
                                       \label{eq:1}
\end{eqnarray}
where $F^{(q)}$, $A^{(q-1)}$ and $B^{(q+1)}$ represent physical fields of 
differential forms (the upper suffices mean the ranks of differential forms)
with $m$ and $\mu$, masses of the latter two, and $G^{(q)}$, $J^{(q-1)}$ and
$K^{(q+1)}$ represent external sources.
Here $(A,B)$ (inner product)$ = \int A \wedge \ast B$, and $d$ and $\delta$ are
conventional {\it boundary operators}.
We assume, 
for convenience, that our spacetime is Euclidian of $d$ dimensions 
\cite{Miyazaki}.

Functionally differentiating the parent action with respect to the field 
$F^{(q)}$ and setting the result to vanish, we express the auxiliary
field $F^{(q)}$ by $A^{(q-1)}$, $B^{(q+1)}$ and $G^{(q)}$ as follows.
\begin{equation}
F^{(q)} = d A^{(q-1)} + \delta B^{(q+1)} + G^{(q)} .
\label{eq:2}
\end{equation}  
Eliminating $F^{(q)}$ from Eq.(\ref{eq:1}) with this relation, we have 
an action for the fields $A^{(q-1)}$ and $B^{(q+1)}$.
\begin{eqnarray}
S_{p} \rightarrow 
S_{1}[A^{(q-1)},B^{(q+1)}]&=&-\frac{1}{2}(d A^{(q-1)} , d A^{(q-1)})
                             -\frac{1}{2}(\delta B^{(q+1)} , \delta B^{(q+1)})
                             -\frac{1}{2}(G^{(q)} , G^{(q)}) \nonumber \\
                          & &-(d A^{(q-1)} , G^{(q)})-(\delta B^{(q+1)} , G^{(q)}) \nonumber \\
                          & &+(A^{(q-1)} , J^{(q-1)})
                             +(B^{(q+1)} , K^{(q+1)})      \nonumber \\
                          & &+\frac{m^2}{2}(A^{(q-1)} , A^{(q-1)})
                             +\frac{{\mu}^{2}}{2}(B^{(q+1)} , B^{(q+1)}) .
\label{eq:3}
\end{eqnarray}   
Namely, we have a dynamical system for the fields $A^{(q-1)}$ and $B^{(q+1)}$.

From Eq.(\ref{eq:3}) follows the equations of motion for the fields
$A^{(q-1)}$ and $B^{(q+1)}$.
\begin{equation}
\delta d A^{(q-1)} - m^2 A^{(q-1)} + \delta G^{(q)} - J^{(q-1)} = 0 ,
\label{eq:4}
\end{equation}   
\begin{equation}
d \delta B^{(q+1)} - {\mu}^{2} B^{(q+1)} + d G^{(q)} - K^{(q+1)} = 0 .
\label{eq:5}
\end{equation}   
Note that the external field $G^{(q)}$ which couples $F^{(q)}$ enters in both 
equations (\ref{eq:4}) and (\ref{eq:5}).
If the spacetime is {\it flat} and compact, the operators $ \delta d$ and
$d \delta$ reduce to the $d$-dimensional Laplacian.
In this case, putting $q=2$ with vanishing external fields in Eq.(\ref{eq:4}),
we have the equation of motion for the free massive gauge field.
Equation(\ref{eq:5}), on the contrary, is the one for a new kind of ($q+1$)-
form field $B^{(q+1)}$.
In the case of the spacetime {\it not} being flat, the 
metric$(g^{\mu \nu})$ which shows ``curvedness'' of spacetime 
appears many times when one expresses $\delta d A^{(q-1)}$ and 
$d \delta B^{(q+1)}$ in {\it component fields}.
Putting $ m = \mu = 0 $ makes Eqs.(\ref{eq:4}) and (\ref{eq:5}) reduce to the 
ones developed in Ref.\cite{Miyazaki} by the present authors.
Specifically, when we take $q=2 , m=0 , G^{(2)} =0$ and the spacetime being 
flat, we have the well-known Maxwell equation with $J^{\mu} (J^{(1)} = 
J_{\mu}d x^{\mu})$, an electric current.
Without those restrictions, the equation
\begin{equation}
\delta d A^{(q-1)} = J^{(q-1)} 
\label{eq:6}
\end{equation}   
is the Maxwell equation in the {\it curved spacetime}.
Equation(\ref{eq:5}) with vanishing $\mu$ and $G^{(q)}$ represents the 
dynamical system with a monopole current $K^{(q+1)}$ \cite{Miyazaki}.

Next let us regard the fields $A^{(q-1)}$ and $B^{(q+1)}$ as independent field variables, and, functionally differentiating the parent action (\ref{eq:1})
with respect to these variables, we have the following equations of motion.
\begin{equation}
m^2 A^{(q-1)} = \delta F^{(q)} -J^{(q-1)} ,
\label{eq:7}
\end{equation}
\begin{equation}
{\mu}^{2} B^{(q+1)} = d F^{(q)} - K^{(q+1)} .
\label{eq:8}
\end{equation}   
With these equations we eliminate $A^{(q-1)}$ and $B^{(q+1)}$ from the parent
action, and we have an action for the field $F^{(q)}$.
\begin{eqnarray}
S_{p} \rightarrow S_{2}[F^{(q)}]&=&\frac{1}{2}(F^{(q)} , F^{(q)})
                                   -(F^{(q)} , G^{(q)}) \nonumber \\
                                & &-\frac{1}{2 m^2}(\delta F^{(q)} - J^{(q)} , 
                                                    \delta F^{(q)} - J^{(q)}) \nonumber \\
                                & &-\frac{1}{2 {\mu}^{2}}(d F^{(q)} - K^{(q+1)} , 
                                                          d F^{(q)} - K^{(q+1)}) .
\label{eq:9}
\end{eqnarray}  
The action(\ref{eq:9}) gives us the equation of motion for the field $F^{(q)}$:
\begin{equation}
(\frac{1}{m^2} d \delta + \frac{1}{{\mu}^{2}}\delta d) F^{(q)} - F^{(q)} 
+ G^{(q)}-\frac{1}{m^2} d J^{(q-1)} -\frac{1}{{\mu}^2} \delta K^{(q+1)} = 0 ,
\label{eq:10}
\end{equation}   
whose special case of the vanishing external fields with $m = \mu $ represents 
nothing but the generalized Klein-Gordon equation for the $q$-form field
$F^{(q)}$ with mass $m$.

The actions(\ref{eq:3}) and (\ref{eq:9}) follow from the one and the same 
parent action (\ref{eq:1}) and in this sense they are dual to each other
\cite{Lindstrom}:
the physical worlds that both actions visualize are dual.
With Eqs.(\ref{eq:7}) and (\ref{eq:8}) the equations of motion (\ref{eq:4})
and (\ref{eq:5}) transform into Eq.(\ref{eq:10}).

Note, here, that the {\it gauge transformations} 
\begin{equation}
A^{(q-1)} \rightarrow A^{(q-1)} + d C^{(q-2)}
\label{eq:11}
\end{equation}   
and 
\begin{equation}
B^{(q+1)} \rightarrow B^{(q+1)} + \delta D^{(q+2)} ,
\label{eq:12}
\end{equation}   
with $C^{(q-2)}$ and $D^{(q+2)}$, arbitrary ($q\!-\!2$)- and ($q\!+\!2)$- 
forms,
resp., make the actions {\it invariant except the mass terms}. 
The gauge transformation (\ref{eq:12}) looks a little peculiar,
because the differential higher than $B^{(q+1)}$ by order one enters in 
$\delta D^{(q+2)}$. 
It plays an essential role in Eq.(\ref{eq:5}) for $\mu =0$ 
[see Ref.\cite{Miyazaki} in this context].\\

Up to now, we have considered the fields $A^{(q-1)} , B^{(q+1)}$ and $F^{(q)}$
as fundamental and, starting with one and the same parent action, obtained the 
dual actions, one of which gives us the generalized Maxwell equation with 
a magnetic monopole.
However, if the fields $A^{(q-1)}$ and $B^{(q+1)}$ are massive, 
the {gauge\it-noninvariance} comes from the mass terms, 
just in the same way as in the Proca equation of the massive spin 1 particle.

From now on we go along an alternative way, i.e., our guiding principle should
be {\it gauge invariance}.
The parent action with which to start is
\begin{eqnarray}
S_{p}&=&\frac{1}{2}(F^{(q)} , F^{(q)})-(F^{(q)} , d A^{(q-1)})
        -(F^{(q)} , \delta B^{(q+1)}) \nonumber \\
     & &+(A^{(q-1)} , \delta J^{(q)})+(B^{(q+1)} , d K^{(q)}) .
\label{eq:13}
\end{eqnarray}   

And, {\it $\grave{a}$ la} Smailagic \cite{Smailagic}, the $q$-form $J^{(q)}$ is 
to be of a special type of the following form:
\begin{equation}
J^{(q)}=m \ast A^{(q-1)} , 
\label{eq:14}
\end{equation}     
where, in order that Hodge's star operator($\ast$) applying to the ($q\!-\!1$)-
form $A^{(q-1)}$ should give the $q$-form $J^{(q)}$, the spacetime dimension 
$d$ is related to $q$ by the relation $d=2q-1$. 
In addition, so as {\it not to have a vanishing mass term}, one further 
restricts the relation between the dimension and the rank as follows. 
\begin{equation}
d=4n-1, ~ q=2n ~ (n=1,2,3,\cdots) . 
\label{eq:15}
\end{equation}   
Then we have, as a mass term, 
\begin{equation}
(A^{(q-1)} , \delta  J^{(q)}) 
= m \int A^{(q-1)} \wedge d A^{(q-1)} ,
\label{eq:16}
\end{equation}
which is nothing but the Chern-Simons term, and invariant under the gauge 
transformation (\ref{eq:11}) up to the total derivative. 
This is equal to that discussed by Deser, 
Townsend et al. \cite{Townsend, Deser} for the flat spacetime in case of 
$q=2$ and $d=3$.

In the same way, for the field $B^{(q+1)}$ to have mass $\mu$ , keeping gauge 
invariance, we choose the field relation
\begin{equation}
K^{(q)}=\mu \ast B^{(q+1)} ,
\label{eq:17}
\end{equation}
as well as the relation between the spacetime dimension and the field rank
\begin{equation}
d=4n-1, ~ q=2n-1 ~ (n=1,2,3,\cdots) . 
\label{eq:18}
\end{equation}   
The mass term becomes 
\begin{equation}
(B^{(q+1)} , d K^{(q)})
= \mu \int B^{(q+1)} \wedge \delta B^{(q+1)} ,
\label{eq:19} 
\end{equation}  
which is {\it invariant under the transformation} (\ref{eq:12}) up to the 
surface integral.
We will call the form (\ref{eq:19}) the {\it generalized} Chern-Simons 
term. \\

Based on the fact that we have mentioned above, we are to have two 
attractive actions which represent massive fields {\it gauge-invariantly}.
The spacetime dimensions $d$ are restricted to be of the form 
$d=4n-1 ~ (n=1,2,3,\cdots)$ and $q$ is given by $q=(d+1)/2$.

\noindent
{\bf Case 1}: {\it let the parent action be given by the following form}:
\begin{equation}
S_{p}[F^{(q)},A^{(q-1)}]=\frac{1}{2}(F^{(q)} , F^{(q)})-(F^{(q)} , d A^{(q-1)})
+m(A^{(q-1)} ,\ast d A^{(q-1)}) .
\label{eq:20}
\end{equation}   
Differentiate functionally Eq.(\ref{eq:22}) with respect to the field $F^{(q)}$,
we have 
\begin{equation}
F^{(q)} = d A^{(q-1)} . 
\label{eq:21}
\end{equation}   
Eliminating $F^{(q)}$ from the parent action with this relation,
we obtain the action of the field $A^{(q-1)}$.
\begin{equation}
S_{p} \rightarrow 
S_{1}[A^{(q-1)}]=-\frac{1}{2}(d A^{(q-1)} , d A^{(q-1)})
+m(A^{(q-1)} ,\ast d A^{(q-1)}) .
\label{eq:22}
\end{equation}
This is an equivalent action to that of Deser et al. \cite{Deser}, generalized
to our case, i.e. , for the non-flat spacetime with odd high dimensions.
The equation of motion for $A^{(q-1)}$ due to (\ref{eq:22}) is
\begin{equation}
\delta d A^{(q-1)} - 2m \ast d A^{(q-1)} = 0 ,
\label{eq:23}
\end{equation}   
which is manifestly gauge-invariant under (\ref{eq:11}).

We now first regard the field $A^{(q-1)}$ as independent and obtain the 
equation of motion from the parent action (\ref{eq:20}):
\begin{equation}
\delta F^{(q)} - 2m \delta \ast A^{(q-1)} = 0 ,
\label{eq:24}
\end{equation}   
the solution of which is given, up to the ambiguity of adding 
$d \ast H^{(q+1)}$ ($H^{(q+1)}$ : an arbitrary ($q\!+\!1$)- form), by
\begin{equation}
A^{(q-1)} = \frac{1}{2m} \ast F^{(q)} .
\label{eq:25}
\end{equation}   
Then, as usual, we have an action for the field $F^{(q)}$
\begin{equation}
S_{p} \rightarrow 
S_{2}[F^{(q)}]=\frac{1}{2}(F^{(q)} , F^{(q)})
-\frac{1}{4m}(F^{(q)} , d \ast F^{(q)}) ,
\label{eq:26}
\end{equation}   
as well as the equation of motion for the field $F^{(q)}$
\begin{equation}
F^{(q)} - \frac{1}{2m} d \ast F^{(q)} = 0 .
\label{eq:27}
\end{equation}  

\vspace*{5ex}
\noindent
{\bf Case 2}: {\it let the parent action be given by the following form}:
\begin{equation}
S_{p}[F^{(q-1)},B^{(q)}]=\frac{1}{2}(F^{(q-1)} , F^{(q-1)})
-(F^{(q-1)} , \delta B^{(q)})
+ \mu (\delta B^{(q)} , \ast B^{(q)}) .
\label{eq:28}
\end{equation}   
After functionally differentiating Eq.(\ref{eq:28}) with respect to the field
$F^{(q-1)}$, obtaining the equation of motion for it,
\begin{equation}
F^{(q-1)} = \delta B^{(q)} ,
\label{eq:29}
\end{equation}  
we have an action for the field $B^{(q)}$.
\begin{equation}
S_{p} \rightarrow 
S_{1}[B^{(q)}]=-\frac{1}{2}(\delta B^{(q)} , \delta B^{(q)})
+\mu (\delta B^{(q)} ,\ast B^{(q)}) .
\label{eq:30}
\end{equation}
Therefore, the equation of motion for the field $B^{(q)}$ becomes
\begin{equation}
d \delta B^{(q)} - 2 \mu d \ast B^{(q)} = 0 ,
\label{eq:31}
\end{equation}   
or, equivalently,
\begin{equation}
(\ast d - 2 \mu ) \delta B^{(q)} = 0 ,
\label{eq:32} 
\end{equation} 
which manifestly shows the invariance under the gauge transformation 
(\ref{eq:12}).

Going along the same line we have the equation of motion for the field 
$B^{(q)}$ through the parent action (\ref{eq:28}):
\begin{equation}
d F^{(q-1)} - 2 \mu d \ast B^{(q)} = 0 ,
\label{eq:33}
\end{equation}  
whose solution is given, up to the ambiguity of adding 
$\ast dE^{(q-2)}$ ($E^{(q-2)}$ : an arbitrary $(q\!-\!2$)-form), by
\begin{equation}
B^{(q)}=\frac{1}{2 \mu } \ast F^{(q-1)} .
\label{eq:34}
\end{equation}   
Hence, eliminating $B^{(q)}$ from the parent action, we obtain the action 
for $F^{(q-1)}$.
\begin{equation}
S_{p} \rightarrow
S_{2}[F^{(q-1)}]=\frac{1}{2}(F^{(q-1)} , F^{(q-1)})
-\frac{1}{4 \mu}(F^{(q-1)} , \ast d F^{(q-1)}) ,
\label{eq:35}
\end{equation} 
from which we finally have the equation of motion for the field $F^{(q-1)}$:
\begin{equation}
F^{(q-1)}-\frac{1}{2 \mu } \ast d F^{(q-1)} = 0.
\label{eq:36}
\end{equation}  

\vspace*{5ex}
Now we mention our concluding remarks. 
We have discussed, based on the parent actions (\ref{eq:1}), (\ref{eq:13}) and (\ref{eq:20}), the dual relations between actions, and also between equations of motion. 
If one wants to have the formulation containing the case where the Maxwell 
equation with an electric charge and a magnetic monopole plays an important 
role, then one is to start with (\ref{eq:1}).
But the mass terms, if they should exist, violate the invariance under the
gauge transformations. 
If one wants to adopt mass terms {\it gauge-invariantly}, one should begin 
with (\ref{eq:13}) or (\ref{eq:20}).
One of the advantages of the method in the present paper is that we formulate 
in a general differential form, not writing down the fields in components.
The component-field method on {\it curved} spacetime of high dimensions 
makes us great efforts in actual calculations.
The results, however, are beautiful likewise.
In this viewpoint see the various examples of Ref.\cite{Miyazaki}.
Any way, our general method gives us a clear meaning and manipulation of a new 
kind of gauge transformations (\ref{eq:12}).
As a matter of fact, every field is defined over a {\it curved} 
high-dimensional Riemannian space. 

\vspace{5ex}
One of the authors(H.E.) would like to thank Iwanami F\=ujukai for financial 
support.

\end{document}